# THERMAL EFFECTS FOR SHAFT–PRE-STRESS ON ROTOR DYNAMIC SYSTEM

**Hisham AL-KHAZALI[1]\* & Mohamad ASKARI[2]**
[1]\* Faculty of Science, Engineering and Computing, Kingston University London,
School of Mechanical & Automotive Engineering,  London, SW15 3DW, UK
K0903888@kingston.ac.uk
[2] Faculty of Science, Engineering and Computing, Kingston University London,
School of Aerospace & Aircraft Engineering, London, SW15 3DW, UK
M.Askari@Kingston.ac.uk

**ABSTRACT**
This Paper outlines study behaviour of rotating shaft with high speed under thermal effects. The method of obtaining the frequency response functions of a rotor system with study whirl effect in this revision the raw data obtained from the experimental results (using Smart Office program) are curve-fitted by theoretical data regenerated from some of the experimental data and simulating it using finite element (ANSYS 12). (FE) models using the Eigen analysis capability were used to simulate the vibration. The results were compared with experimental data show analysis data with acceptable accuracy and performance.
The rotating effect causes un-symmetry in the system matrices, resulting in complexity in decoupling the mathematical models of the system for the purpose of modal analysis. Different method is therefore required, which can handle general system matrices rather than symmetrical matrices, which is normal for passive structures. Mathematical model of the system from the test data can be assembled. The frequency response functions are extracted, Campbell diagram are draw and simulated. (FE) is used to carry out such as simulation since it has good capability for Eigen analysis and also good graphical facility.

## 1.　INTRODUCTION

Rotor dynamics is the key in design and maintenance of all rotating machinery, which forms the backbone of industrial infrastructure and plays an important role in the growth and development of the country**.** Most of the research investigations are performed based on vibration characteristics of rotor dynamics system. Nelson and Mc Vaugh [1] first used the finite element method and derived various property matrices. They gave the detailed information about critical speed and unbalance response. Instability due to internal damping and hysteretic damping is given by Zozi and Nelson [2]. Lin and Lin [3] have investigated the optimal weight design of rotor-bearing system. Aleyaasin et al., [4] have used the dynamic stiffness matrix method to perform the lateral vibration analysis of rotor-dynamic system. Firoozian and Zhu [5] have used transfer matrix method to study lateral vibration of rotor-bearing system. Wu and Yang [6] used transfer matrix method, finite element method and Lagrangian dynamics method to study coupled tensional lateral vibration analysis of rotor-bearing system. All the literature contains research carried out on actual (Full-size) rotor bearing system. Wu [7] developed a methodology to study vibration characteristics of full sized rotor system based on scaled models (Scaled-Rotor). He gave a systematic approach to the methodology of scaling laws and dimensional analysis applied to a rotor dynamic system. The theory and predicted values agree with each other in case of Eigen frequencies and steady state response.

## 2.　CAMPBELL DIAGRAM COMPARISON INCLUDES PRE-STRESS DUE TO THERMAL LOADING:

### 2.1. Whirl Effect
A gyroscope is a device for measuring or maintaining orientation, based on the principles of angular momentum [8, 9]. In essence, a mechanical gyroscope is a spinning wheel or disc whose axle is free to take any orientation. Although this orientation does not remain fixed, it changes in response to an external torque much less and in a different direction than it would without the large angular momentum associated with the disk's high rate of spin and moment of inertia. Since external torque is minimized by mounting the device in gimbals, its orientation remains nearly fixed, regardless of any motion of the platform on which it is mounted. The resulting gyroscopic matrix,[G]. Gyroscopic effect occurs whenever the mode shape has an angular (conical /rocking) component. Also at the same section it is mentioned that having an overhung load will make the gyroscopic occurs in both, $1^{st}$ and $2^{nd}$ mode of vibration that mean gyroscopic effect occurs at the lower frequencies. Although it is sufficient to allocate the experiment in the $1^{st}$ mode,[8, 10]. Due to gyroscope effect produced (Whirl movement), when a rotating structure vibrates at its resonant frequency, points on the spin axis undergo an orbital motion, called whirling. Whirl motion can be a forward whirl (**FW**) if it is in the same direction as the rotational velocity or backward whirl (**BW**) if it is in the opposite direction [11, 12& 13].

## 2. 2.Theoretical Investigation

### 2.2.1. Mathematical models for active structure (rotating structures)

The general equations of motion for a rotating and supporting structure are given by [14]:

$$[M]\{\ddot{q}(t)\}+[(C+G)(\Omega)]\{\dot{q}(t)\}+[K(\Omega)]\{q(t)\}=\{F(t)\} \quad (1)$$

Where [c] includes the gyroscopic effect.

$$C \neq C^T \qquad K \neq K^T \quad (2)$$

A trial solution of cyclic of the form, and the Eigen vectors are obtained by solving

$$[s_r^2[M]+s_r[C]+[K]]\{\psi R\}_r = \{0\} \quad (a)$$

$$[s_r^2[M]^T+s_r[C]^T+[K]^T]\{\psi L\}_r = \{0\} \quad (b) \quad (3)$$

Where $s_r$, $\{\Psi R\}_r$ and $\{\Psi L\}_r$ are the complex Eigen values, right Eigen vectors and left Eigen vectors respectively for mode $r$.

$$\{u(t)\} = \begin{Bmatrix}\{q(t)\}\\\{\dot{q}(t)\}\end{Bmatrix} \qquad \text{State space} \quad (4)$$

Combining the above equations:

$$\begin{bmatrix}[C] & [M]\\[M] & [0]\end{bmatrix}\{\dot{u}(t)\}+\begin{bmatrix}[K] & [0]\\[0] & -[M]\end{bmatrix}\{u(t)\}=\begin{Bmatrix}\{F(t)\}\\\{0\}\end{Bmatrix} \quad (5)$$

Or

$$[A]\{\dot{u}(t)\}+[B]\{u(t)\}=\begin{Bmatrix}F(t)\\0\end{Bmatrix} \quad (6)$$

$$\{q(t)\}=c_1\{\psi R\}_1 e^{s_1 t}+c_2\{\psi R\}_2 e^{s_2 t}+\ldots\ldots +c_{2n}\{\psi R\}_{2n}e^{s_{2n}t} \quad (a)$$

$$\equiv \{q(t)\}=\sum_{r=1}^{2n} c_r \{\psi R\}_r e^{s_r t} \qquad (b) \quad \text{Vector space} \quad (7)$$

$$\{q(t)\}=\sum_{r=1}^{2n} d_r \{\psi L\}_r e^{s_r t} \qquad (c)$$

$$\{u(t)\}=\begin{Bmatrix}q(t)\\\dot{q}(t)\end{Bmatrix}=\begin{Bmatrix}\bar{q}\\S\bar{q}\end{Bmatrix}e^{st} \quad (8)$$

$$\{\dot{u}(t)\}=\begin{Bmatrix}\dot{q}(t)\\\ddot{q}(t)\end{Bmatrix}=\begin{Bmatrix}S\bar{q}\\S^2\bar{q}\end{Bmatrix}e^{st} \qquad \text{State space} \quad (9)$$

For simplification the following terms are introduced:

$$_rG_{jk}={_r\phi R_j}\,{_r\phi L_k} \qquad \text{and} \qquad {_rG_{jk}}^*={_r\phi R_j}^*\,{_r\phi L_k}^* \quad (10)$$

Where $\emptyset R$ and $\emptyset L$ are normalised Eigen vectors.

Bi-Orthogonality relationships of left and right Eigen vector could be used to diagonals the reacceptance matrix and cross reacceptance between two coordinate's j and k may be obtained as follows [10, 12& 14]:

$$H_{jk}(\omega)=\sum_{r=1}^{n}\left(\frac{{_r\phi R_j}\,{_r\phi L_k}}{(i\omega-s_r)}+\frac{{_r\phi R_j}^*\,{_r\phi L_k}^*}{(i\omega-s_r^*)}\right) \quad (11)$$

However the complex Eigen values have the following forms:

$$s_r, s_r^* = -\omega_r \zeta_r \pm i\omega_r \sqrt{1-\zeta_r^2} \qquad (12)$$

Substituting equation (12) into (11) and simplifying gives:

$$H_{jk}(\omega) = \sum_{r=1}^{n} \left( \frac{2\omega_r \left( \zeta_r \operatorname{Re}(_rG_{jk}) - \sqrt{1-\zeta_r^2} \left( \operatorname{Im}(_rG_{jk}) \right) \right) + i\left( 2\omega \operatorname{Re}(_rG_{jk}) \right)}{\left( \omega_r^2 - \omega^2 + 2i\omega\omega_r\zeta_r \right)} \right) \qquad (13)$$

Equation (13) represents the reacceptance between two coordinates j and k for a system with n degree of freedom. The denominator is identical to the denominator of the reacceptance expression for an n degree of freedom system with symmetric system matrices, but the numerator is very different.

## 2.3. Experimental Setup

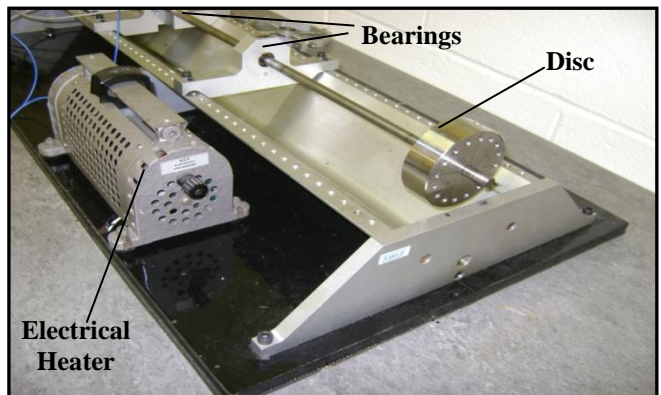

*Picture 1, Experimental setup for thermal gyroscopic modal testing.*

The test rotor is shown in picture 1. Basically; the rotor consisted of a shaft with a nominal diameter of 10 mm, with an overall length of 610 mm. Two plain bearings, RK4 Rotor Kit made by Bentley Nevada (the advanced power systems energy services company), are used to extract the necessary information for diagnostic of rotating machinery, such as turbines and compressor. The testing of the process will be conducted on the rotary machine as the project is based on rotary dynamics reach practical results for the purpose of subsequently applied machinery rotary by using (Smart Office program), and then do the experimental testing using the impact test, installed fix two accelerometer (model 333B32), sensitivity(97.2&98.6) mv/g in Y&Z direction and roving the hammer (model 4.799.375, S.N24492), on each point for the purpose of generating strength of the movement for the vibration body and the creation of vibration for that with creating a computer when taking readings in file that was dimensions and introducing it with the data within the program (Smart Office)[15, 16& 17]. Configurations of testing on the rotary machines, all necessary equipment for test were shown with the geometry design wizard (see Fig. 1).

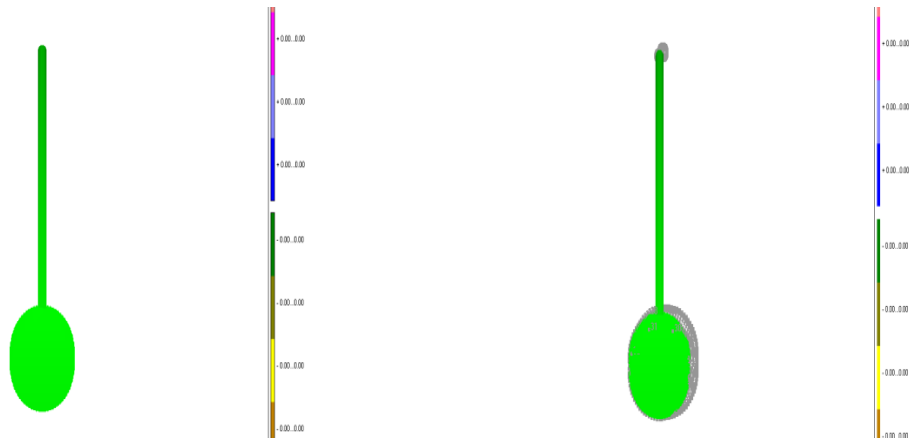

*Figure 1, Geometry design for modal (gyroscopic effect), experimental test using smart office.*

*Table 1.*
*Backward & Forward natural frequencies in experimental thermal effect on shaft–pre-stress of gyroscopic effect for $1^{st}$ mode shape.*

| Speed of Rotation(RPM) | Experimental Results(Hz),BW | Experimental Results(Hz),FW |
| --- | --- | --- |
| 30 | 18.537 | 18.537 |
| 2000 | 17.467 | 18.986 |
| 3000 | 17.184 | 19.125 |
| 4000 | 17.095 | 19.425 |
| 6000 | 16.995 | 19.713 |

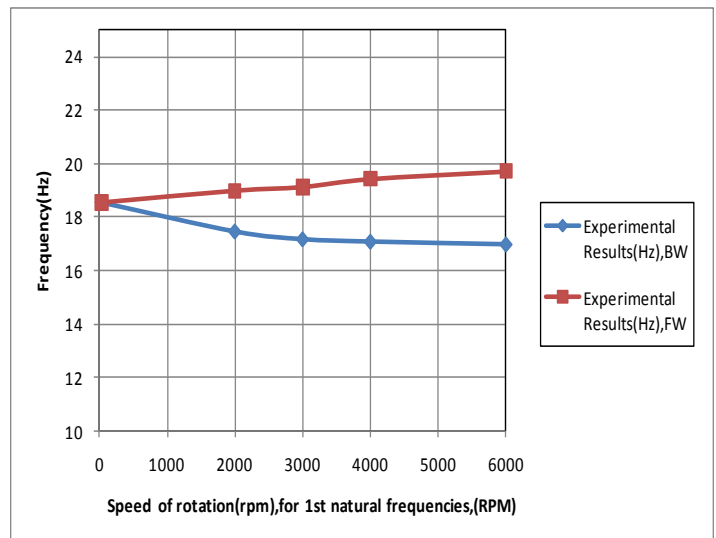

*Figure 2, Experimental thermal calculations on gyroscopic effect.*

This introduces the establishment and simplification of the temperature field and the general calculation method of temperature stress of the pre-stressed.

### 2.4. Simulation of Model (Finite Element)
### 2.4.1. Simulation of a simple model in (ANSYS 12.)
A simple model of rotor system with an overhung disc with multi degree of freedom ( Y and Z directions) has been used to demonstrate the above capability (see Fig. 3)[18]. A program has been written in (ANSYS 12), Applying of gyroscopic effect to rotating structure was carried by using (CORIOLIS) command. [2,19& 20].

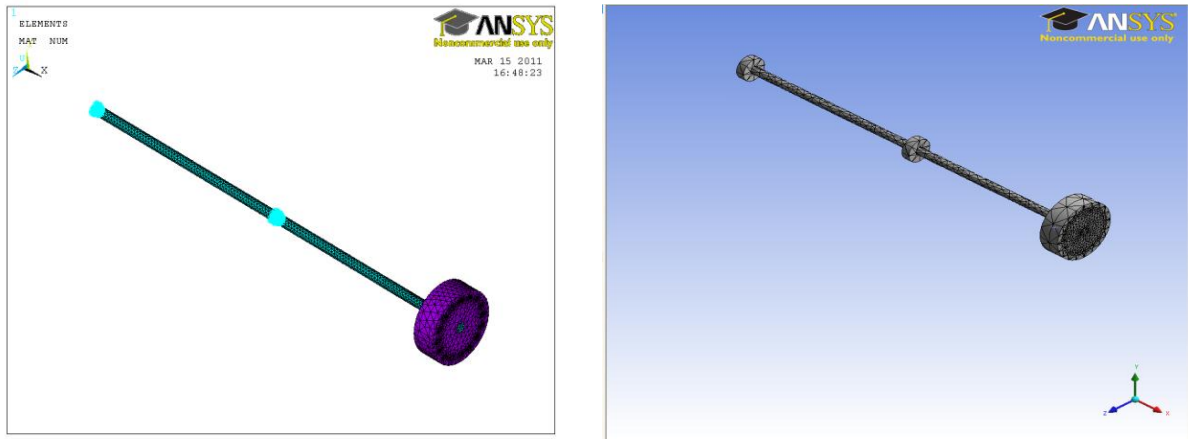

*Figure 3, A-Finite element model (Gyroscopic Geometry) in ANSYS (APDL), B-(ANSYS) work bench.*

### 2.4. 2. Validate simulation results under thermal effect
Like it was discussed, this fact is known as gyroscopic effect and occurs because of the Backward (BW) and forward (FW) whirl of the rotor. The Eigen vectors would show these phenomena,[12, 21]. As the speed increases the gap between the two frequencies becomes larger (See Fig 4- A, B).

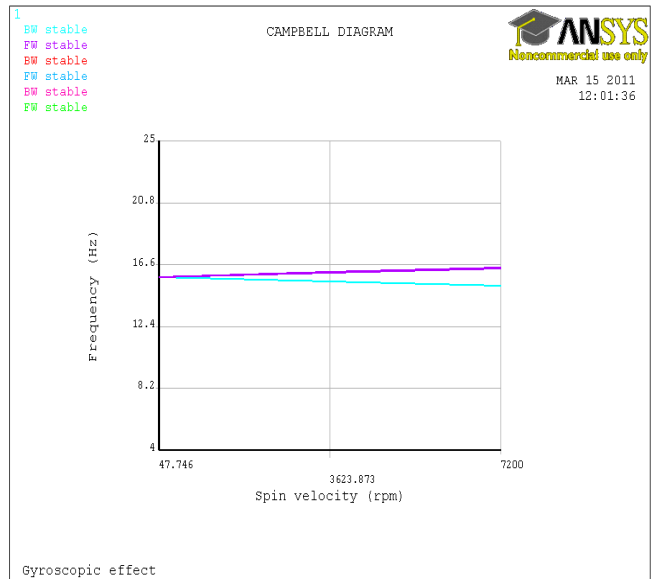

*A, No Pre-Stress.*

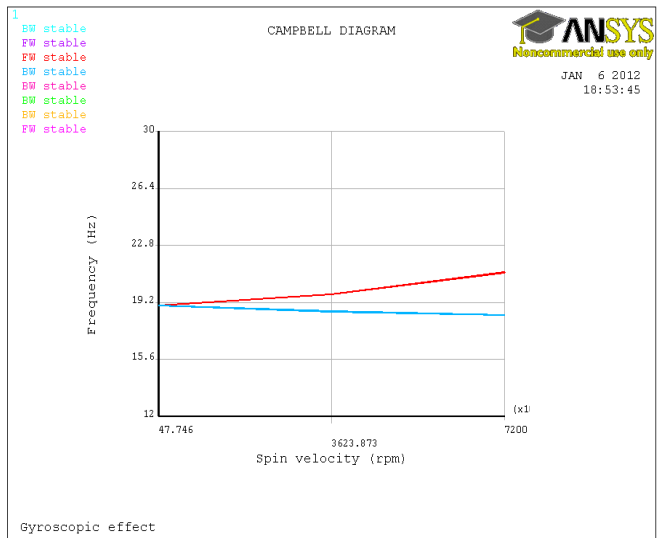

*B, With thermal pre-stress.*
*Figure 4, Thermal body load up to 1500 Deg F;*

### 3.   OVERALL DISCUSSION AND CONCLUSION
We can see from Table 1, and Fig. 2, Experimental thermal calculations on gyroscopic effect of thermal body load up to 1500 deg F, and we can comparison with finite element analysis using ANSYS simulation Fig.(4-B), when heating the shaft until 1500 deg F, the frequency is goes up and become 19 Hz, While still approximately 15 with no pre-stress without heating see Fig.(4-A). That mean due to rising temperature was effect on mechanical properties (structural characteristics) of shaft and then due to increasing the value of natural frequency at the same condition, because a vibrating structure by nature deforms into certain patterns at the natural frequencies of the structure. The deformation patterns and their natural frequencies depend only on the characteristics of the structure, and not on the input force.

## 4. ACKNOWLEDGEMENTS
The authors are gratefully acknowledging support of the Iraqi Ministry of Higher Education, Iraqi Cultural Attaché in London and Kingston University London for Completion this research.